**Large anomalous Nernst and inverse spin-Hall effects in epitaxial thin films of Kagome semimetal Mn$_3$Ge**

Deshun Hong[1], Naween Anand[1], Changjiang Liu[1], Haihua Liu[2], Ilke Arslan[2], John E. Pearson[1], Anand Bhattacharya[1] and J. S. Jiang[1]*

[1]*Materials Science Division, Argonne National Laboratory, Argonne, Illinois 60439, USA;*

[2]*Center for Nanoscale Materials, Argonne National Laboratory, Argonne, Illinois 60439, USA;*

Synthesis of crystallographically well-defined thin films of topological materials is important for unraveling their mesoscale quantum properties and for device applications. Mn$_3$Ge, an antiferromagnetic Weyl semimetal with a chiral magnetic structure on a Kagome lattice, is expected to have enhanced Berry curvature around Weyl nodes near the Fermi energy, leading to large anomalous Hall / Nernst effects and a large spin-Hall effect. Using magnetron sputtering, we have grown epitaxial thin films of hexagonal D0$_{19}$ Mn$_3$Ge that are flat and continuous. Large anomalous Nernst and inverse spin-Hall effects are observed in thermoelectric and spin-pumping devices. The anomalous Nernst signal in our Mn$_3$Ge films is estimated to be 0.1 μV / K, and is comparable to that in ferromagnetic Fe, despite Mn$_3$Ge having a weak magnetization of ~ 3.5 mμ$_B$ / Mn at room temperature. The spin mixing conductance is 90.5 nm$^{-2}$ at the Py / Mn$_3$Ge interface, and the spin-Hall angle in Mn$_3$Ge is estimated to be about 8 times of that in Pt.

## I. INTRODUCTION

Recent discoveries of topological properties in layered Kagome semimetals such as Mn$_3$X (X = Sn [1-5], Ge [4 - 6]), Fe$_3$Sn$_2$ [7 - 9] and Co$_3$Sn$_2$S$_2$ [10, 11] have attracted broad interest in these materials due to their rich physics and potential for applications. For example, as a non-collinear antiferromagnetic material, Mn$_3$Sn has a glide mirror plane where the two Kagome layers in each unit cell can be transformed into one another by an extra c/2 translation. Weyl nodes of opposite chirality are protected by this symmetry and can further give rise to a non-zero Berry flux; the effects are similar to applying a fictitious field of ~ 100 T [2, 4, 12]. With the external magnetic field applied perpendicularly to the glide mirror plane, giant anomalies in the Hall resistance and the Nernst signal which are comparable to those of ferromagnetic materials have been reported [1, 3]. Theoretical calculations also indicate large spin-Hall angles in these materials [13], which are promising for antiferromagnetic spintronics applications. Recently, an magnetic inverse spin-Hall effect (ISHE) [14] as well as a large magneto-optical Kerr effect [15] have been observed in Mn$_3$Sn single crystals.

Compared with Mn$_3$Sn, the hexagonal D0$_{19}$ phase of Mn$_3$Ge has a layered Kagome structure and is antiferromagnetically ordered with a 120° triangular magnetic structure all the way to low temperatures. According to calculations, there are 50 Weyl points in Mn$_3$Ge near the Fermi surface and this number is larger than that in Mn$_3$Sn, presumably due to the lower spin-orbit coupling [4]. Mn$_3$Ge is predicted to have a larger anomalous Hall conductivity than Mn$_3$Sn [13], and its topological properties can persist down to low temperatures while Mn$_3$Sn becomes a spin glass below 50 K [1]. The topological properties of Mn$_3$Ge have been investigated via transport measurements on bulk single-crystal samples [6, 16]. On the other hand, thin films offer the possibility of tuning topological phases with strain, proximity effects and gating, and are well suited for exploration of fundamental mesoscopic transport properties and device applications. Therefore, synthesis of crystallographically well-defined Mn$_3$Ge thin film is of great interest.

Until now, the only D0$_{19}$ Mn$_3$Ge thin films that have been synthesized are polycrystalline, and it is not known whether they exhibit topological properties [17]. In this work, we grow Mn$_3$Ge thin films epitaxially on Ru-buffered single-crystal sapphire by magnetron sputtering. Microstructural analyses indicate the formation of hexagonal D0$_{19}$ Mn$_3$Ge with the c-axis oriented out-of-plane. The Mn$_3$Ge films have a weak ferromagnetic magnetization of ~ 3.5 m$\mu_B$ / Mn at room temperature. A large anomalous Nernst effect (ANE) of 0.1 μV / K is measured, and the high magnetic field (~ 2 T) needed to reverse its sign suggests pinning of antiferromagnetic domains by defects and grain boundaries. The Mn$_3$Ge thin films exhibit high spin-charge conversion efficiency. The measured spin Hall angle is about 8 times of that of the archetypal spin Hall material Pt. The large ANE and spin Hall angle highlight the significance of band topology-induced Berry curvature in Mn$_3$Ge.

## II. EXPERIMENTAL DETAILS

Previous growth of epitaxial Mn$_3$Sn films [18] suggests this high surface energy material tends to have wetting issues and is discontinuous after growth. Here, we find that continuous epitaxial Mn$_3$Ge films can be successfully grown by magnetron sputtering onto Ru-buffered sapphire (0001) substrate at a high growth rate and a lower temperature. Mn and Ge were co-sputtered from elemental sources. Their atomic fluxes were measured *in situ* using a quartz crystal microbalance which had been calibrated using X-ray reflectivity (XRR). To determine the optimal growth conditions, we varied the Mn-to-Ge flux ratio and analyzed the composition of the resulting films using Rutherford backscattering spectrometry (RBS). Remarkably, the formation of (0001)-oriented single-phase D0$_{19}$ Mn$_3$Ge occurred only when the flux ratio was maintained at ~ 5.07, indicating the sticking coefficient of Mn is about 3/5 of the Ge one. The details of the growth are as follows: A 10-nm-thick Ru buffer layer was deposited at 3.6 nm / min at 350°C, and then annealed at 700°C for 15 minutes. This procedure produced a Ru layer which is *c*-axis oriented, and whose surface flatness is evidenced by the presence of sharp streaks in reflection high energy electron diffraction patterns and satellite peaks in high-angle X-ray diffraction (XRD) spectra. The Mn$_3$Ge layer was subsequently deposited onto the Ru template at 400°C and at an Ar pressure of 2.5 mTorr. The effective total deposition rate of the Mn$_3$Ge layer was approximately 10 nm / min. The films were post-annealed in vacuum at 500°C for up to 2 hours to improve crystallinity and chemical order. The heating and cooling rates during all stages of growth were 50°C / min and 20°C / min, respectively. The composition of the final films, determined using RBS, was Mn$_{3.23\pm0.05}$Ge, and is very similar to that of bulk single crystals (Mn$_{3.22}$Ge) where extra Mn is needed to stabilize the D0$_{19}$ structure [16]. For exchange bias and spin Hall measurements, Mn$_3$Ge/permalloy (Py) bilayer structures were also fabricated, with the Py layer sputter-deposited at 6 nm / min after the Mn$_3$Ge layer had cooled to ambient temperature. All films were not capped before being taken out of high vacuum.

XRD and XRR measurements were performed on a Philips X'PERT-PRO MRD system with a Cu source ($\lambda = 1.5406$ Å). The measured reflectivity data were modelled using the GenX software [19]. Atomic force microscopy (AFM) was performed on a Bruker Dimension Icon in soft tapping mode with a standard tip (radius < 10 nm). Micrographs with a 5×5 μm$^2$ scan area were acquired at several locations on each film, which is 10×10 mm$^2$ in size. High resolution transmission electron microscopy (HRTEM) images were taken on a FEI Tecnai F20ST(S)TEM under 200 keV of beam energy. The HRTEM samples were prepared via focused ion beam milling, and further thinned on a low-energy (500 V) Ar ion mill. The magnetic properties of both Mn$_3$Ge and Mn$_3$Ge/Py samples were measured in a Quantum Design MPMS SQUID magnetometer at various temperatures with the magnetic field parallel to the film surface. For exchange bias measurements, the Mn$_3$Ge/Py samples were always cooled from room temperature in a 1-T in-plane magnetic field.

Thermoelectric and spin-pumping devices were fabricated using optical lithography and Ar ion milling. For each thermoelectric device (shown schematically in Fig. 4(a)), a 20×800 μm$^2$ stripe was first patterned from the sputtered Ru/Mn$_3$Ge film. It was then covered with a 100 nm-thick SiN layer grown via chemical vapor deposition for electrical isolation, and followed by a sputtered 50 nm Au layer that would function as the on-chip heater. A reference device, where the Ru/Mn$_3$Ge film was replaced with a 60-nm layer of Fe, was also identically microfabricated. The Nernst effect in the devices was measured at room temperature using the lock-in technique on a Quantum Design Physical Property Measurement System. A sinusoidal current ($f$ = 3 Hz, $I_{pp}$ = 3 mA) applied to the Au heater created an out-of-plane temperature gradient. With an in-plane magnetic field applied perpendicular to the length of the stripe, the Nernst signal was detected along the length of the device at the second harmonic (2$f$).

For the spin-pumping device (shown in Fig. 5(a)), a 1000×200 μm$^2$ bar was patterned from the Ru/Mn$_3$Ge/Py film stack. Two large Au contact pads were deposited onto the bar. Between the contact pads, an electrically isolated (using SiN) coplanar waveguide (CPW) was fabricated on the bar, along the Mn$_3$Ge [1$\bar{1}$00] direction, and was terminated with a 50 Ω resistive load. At room temperature, microwave excitations with varying frequencies (7 - 18 GHz) and power (12 - 18 dBm) were applied to the CPW to drive magnetization precession in the Py layer. An external magnetic field was applied parallel to the CPW axis and was swept through ferromagnetic resonance. The ISHE signal was extracted from the voltage measured across the contact pads. For comparison, reference devices where the Ru/Mn$_3$Ge/Py stack was replaced with a Pt(10nm)/Py(10nm) bilayer or with a Py(10nm) single layer have also been identically microfabricated and measured.

### III. RESULTS AND DISCUSSION

**1. Structural properties**

Fig. 1(a) shows a representative XRD 2θ-ω scan from a sapphire / Ru (10nm) / Mn$_3$Ge (100nm) sample. We intentionally offset ω by 0.2° from the sapphire (0006) reflection to avoid the strong substrate contribution. Only the Mn$_3$Ge (0002) and (0004) peaks and the Ru (0002) and its satellite peaks can be seen in the entire scan range. The position of the Mn$_3$Ge (0002) peak (2θ = 41.910°) gives $c$ = 4.308Å, which is very close to the bulk value of 4.312Å. Furthermore, using the bulk in-plane lattice constant $a$ = 5.352Å, we were able to locate the Bragg peaks of the Mn$_3$Ge {10$\bar{1}$1} and {11$\bar{2}$2} planes at their respective calculated 2θ and χ positions in the azimuthal XRD ϕ-scans, which are shown in Fig. 1(c). We therefore conclude that the sputtered Mn$_3$Ge films are fully relaxed. The full width at half maximum (FWHM) values of the Mn$_3$Ge (0002) and Mn$_3$Ge (0004) peaks are 0.252° and 0.360°, respectively. Using the Scherrer equation, we estimate that the out-of-plane coherence length of the Mn$_3$Ge crystallites is about 32 nm, which is less than the Mn$_3$Ge layer thickness. If the Mn$_3$Ge layer were fully coherent, the FWHM values would be 0.081° and 0.115° for the Mn$_3$Ge (0002) and Mn$_3$Ge (0004) peaks, respectively. The inset of Fig. 1(a) shows the rocking curve of the Mn$_3$Ge (0002) reflection measured at 2θ= 41.9100°. The sharp central peak is the contribution from the sapphire substrate, whereas the contribution from the Mn$_3$Ge layer has a FWHM of 0.5918°± 0.0012°. There is a finite constant background to the rocking curve, indicating that although good crystalline alignment is achieved between the Mn$_3$Ge $c$-axis and the growth direction, crystallites with random misorientations may be present (See Supplemental Material [20] for details).

The XRR curve of the same sample is displayed in Fig. 1(b), which shows intensity modulations arising from interference as the x-ray beam reflects from the various interfaces in the sample. The blue curve is a simulation using a model that describes the sample stack. A Mn$_3$Ge-oxide (Mn$_3$GeO$_x$) layer was needed as the topmost layer in the model to achieve a reasonably good match with the measured data. In the simulation, thicknesses values of 9.5 nm, 98.3 nm, and 8.2nm were used for Ru, $Mn_3Ge$ and $Mn_3GeO_x$, respectively. The roughness values of the sapphire/Ru, Ru/$Mn_3Ge$, $Mn_3Ge$/$Mn_3GeO_x$ and $Mn_3GeO_x$/air interfaces are 0.2 nm, 0.6 nm, 1.6 nm and 4.5 nm, respectively. The scattering length density (SLD) for sapphire, Ru and $Mn_3Ge$ are fixed at their bulk values, while a value of 6.8 g/cm$^3$ is used for $Mn_3GeO_x$.

To determine the in-plane orientation of the $Mn_3Ge$ films and the epitaxial relationship, azimuthal XRD $\phi$-scans were performed for the Bragg peaks of the $Mn_3Ge$ and Ru layers and the sapphire substrate. Considering the lattice constant *a* of $Mn_3Ge$ is nearly double that of Ru, while the lattice constant *c* of sapphire is nearly triple those of $Mn_3Ge$ and Ru, care was taken to select Bragg peaks that have no other peaks nearby. Shown in Fig. 1(b) are the $\phi$-scans of $Mn_3Ge$ {10$\bar{1}$1} and {11$\bar{2}$2}, Ru {10$\bar{1}$1}, and sapphire {10$\bar{1}$4} planes. The sapphire {10$\bar{1}$4} planes have a three-fold symmetry, while the Ru basal hexagon is rotated by 30° away from that of sapphire. This demonstrates the well-known rotational honeycomb epitaxy between Ru and sapphire [21]. The two $\phi$-scans of $Mn_3Ge$ show that its basal hexagon is exactly aligned with that of the Ru buffer. However, the peaks in the $Mn_3Ge$ {11$\bar{2}$2} $\phi$-scan have broader bases, indicating that the mosaicity is larger in those directions. It is worth noting that the intensity ratio of the $Mn_3Ge$ {10$\bar{1}$1} and {11$\bar{2}$2} peaks is close to the calculated value based on the structure factor [22]. Since atomic disorders in the unit cell lead to deviations from the ideal structure factor, the agreement between the intensity ratios suggest that there is good atomic ordering in the $Mn_3Ge$ films. In Fig. 1(d) we illustrate the stacking arrangement of the sapphire substrate, the Ru and Mn$_3$Ge layers. The epitaxial relationship among them is: sapphire (0001)[10$\bar{1}$0]|| Ru (0001)[11$\bar{2}$0]||Mn$_3$Ge (0001)[11$\bar{2}$0].

Shown in Fig. 2a is the typical AFM image of a 100nm Mn$_3$Ge film. The root mean square roughness is ~ 2.5 nm over the 5×5 μm$^2$ scan area, consistent with the result of XRR modeling. The maximum height variation in the AFM image is far smaller than the Mn$_3$Ge layer thickness, indicating that, at length scales larger than the lateral resolution (10 nm) of AFM, the film is continuous.

The cross-sectional HRTEM micrograph of a Mn$_3$Ge film is displayed in Fig. 2b. The (002) planes of Ru and Mn$_3$Ge are clearly visible and are parallel to the sapphire/Ru interface. The measured interplanar spacings are 0.210 nm and 0.213 nm for Ru and Mn$_3$Ge, respectively. There is no discontinuity in the film at the 10 nm scale. Also visible in the image are Moiré fringes, as shown in areas outlined by red traces. Moiré fringes in TEM images are interference patterns resulting from two overlapping crystallites that have different lattice constants or orientations; the latter case gives fringes that are nearly perpendicular to the atomic planes. From the periodicity and orientation of the Moiré fringes in Fig. 2b, we deduce they are the result of crystal grains that are misoriented by ~ 17.5° from the film normal. The extent of the Moiré fringes indicates the *misoriented* crystallites are about 5-10 nm in size. The presence of these misoriented crystallites is consistent with the background in the Mn$_3$Ge (0002) rocking curve and the broadened bases of the Mn$_3$Ge {11$\bar{2}$2} ϕ-scan peaks. As will be seen later, the defects and grain boundaries also have implications for the reversal of antiferromagnetic domains in the Mn$_3$Ge films.

## 2. Magnetic properties

Although the hexagonal D0$_{19}$ Mn$_3$Ge has a triangular antiferromagnetic structure, it possesses a weak in-plane ferromagnetic moment arising from spin canting toward the local easy axis [23]. In bulk Mn$_3$Ge single crystals, the ferromagnetic moment amounts to 6 - 8 m$\mu_B$/Mn at temperatures between 5 and 300 K, and it is noted that this in-plane moment is essential for controlling chirality of the spin structure via an applied magnetic field [6, 16]. Shown in Fig. 3(a) is the field dependence of the ferromagnetic component of the in-plane magnetization of a 100 nm Mn$_3$Ge film measured at various temperatures. The ferromagnetic component of the magnetic moment was obtained by subtracting a linear background from the raw magnetometry data (See Supplemental Material [20] for details). This ferromagnetic magnetization increases with decreasing temperature. At 300K and H = 2 T, the ferromagnetic moment per Mn atom in the Mn$_3$Ge films is ~ 3.5±0.4 m$\mu_B$, which is close to that of the bulk. The coercive field for switching the ferromagnetic moment is ~ 60 mT at 300 K, decreases slightly to 50 mT at 100K, before increasing to 110 mT at 10 K. In Fig. 3 (b), we present the hysteresis loop of a Mn$_3$Ge(100nm)/Py(10nm) bilayer structure measured at 10 K after cooling from room temperature in a 1-T in-plane magnetic field. The loop is shifted in the negative field direction by ~7.5 mT. Given that the triangular spin structure is not expected to have in-plane magnetic anisotropy, and that the coercivity of the bilayer (H$_C$ = 25.5 mT) is significantly larger than that of a single Py layer (typically less than 1 mT), the occurrence of exchange bias indicates that the antiferromagnetic domains in the Mn$_3$Ge layer are pinned, possibly by defects and grain boundaries.

### 3. Anomalous Nernst effect

As a counterpart of the anomalous Hall effect (AHE) where contributions from Weyl nodes well above the Fermi energy ($E_F$) can be very small, ANE depends on the Berry curvature near $E_F$, involving states both above and below $E_F$ over an extent of energy determined by the broadening of the Fermi function [24]. In $Mn_3Ge$, transport anomalies can be captured only when the external magnetic field is applied in the Kagome plane [6, 12]. The *c*-axis oriented $Mn_3Ge$ films are thus well-suited for integration into a thermoelectric device that employs out-of-plane thermal transport to probe the Berry curvature-driven ANE.

For the measurement geometry shown in Fig. 4(a) where the temperature gradient $\nabla T$ is out-of-plane in the *z*-direction (along $Mn_3Ge$ [0001]) and the applied magnetic field is in-plane in the *y*-direction (along the device width and $Mn_3Ge$ [0$\bar{1}\bar{1}$0]), the ANE signal ($S_{xz}$) is given by:

$$S_{xz} = E/\nabla T = [(\rho_{Mn3Ge}/t_{Mn3Ge} + \rho_{Ru}/t_{Ru})(t_{Ru}/\rho_{Ru})](V_{Nernst}/L)(1/\nabla T), \qquad (1)$$

where $E$ is the electric field created by ANE in the *x*-direction (along the device length and $Mn_3Ge$ [2$\bar{1}\bar{1}$0]), $V_{Nernst}$ is measured anomalous Nernst voltage, and $L$ is the device length. The term in square brackets is the correction factor which accounts for the shunting effect of the Ru underlayer, where $t_{Ru}$ and $t_{Mn3Ge}$ are the layer thicknesses, and $\rho_{Ru}$ and $\rho_{Mn3Ge}$ are the resistivity of the respective layers. Using $t_{Ru}$ =10 nm, $t_{Mn3Ge}$ = 100 nm, $\rho_{Ru}$ = 9 μΩ·cm, and $\rho_{Mn3Ge}$ = 200 μΩ·cm [25], we find the correction factor to be 3.22. $\nabla T$ is given by Fourier's law, $\mathbf{q}=-k\nabla T,$ where $\mathbf{q}$ is the thermal flux density, and $k$ is the thermal conductivity. Form the resistance of the heater at 300 K, we estimate that the ac current applied to the heater generated a peak $q$ of 20.0±0.6 mW/mm². Using $k_{Mn3Ge}$ = 6.8 Wm⁻¹K⁻¹ and $k_{Fe}$ = 83.5 Wm⁻¹K⁻¹ [26, 27], we obtain $\nabla T$ in the Ru/Mn₃Ge and Fe devices as 2.94±0.10 K/mm and 0.24±0.01 K/mm, respectively. The noise level in our $V_{Nernst}$ measurement setup is 4 nV/$\sqrt{Hz}$. Taking into account the uncertainties in the layer thicknesses and the device dimensions, we place the uncertainty in the calculated $S_{xz}$ at about 10%.

Fig. 4(b) presents the field dependence of $S_{xz}$ of a Ru(10nm)/ Mn₃Ge(100nm) device at 300 K. $S_{xz}$ reverses sign when the magnetic field is swept between ±9 T and the reversal is hysteretic. Although $S_{xz}$ starts decreasing as soon as the field polarity is reversed, the sign reversal proceeds very slowly and does not complete until the applied field reaches ~ 7 T, giving a coercivity of ~ 2 T. This is significantly greater than the coercivity of 2 mT to 30 mT for switching AHE in single crystals of Mn₃Ge [6, 16]. The difference is possibly due to the defects and grain boundaries in the Mn₃Ge thin films. Although the chiral antiferromagnetic domain in Mn₃Ge can nucleate reversal easily, the antiferromagnetic domain walls in thin films could become pinned, and increasingly larger fields might then be needed to free the domain walls from pinning sites of various strengths. At 300 K, the saturation $S_{xz}$ = 0.10 μV / K in our Mn₃Ge epitaxial thin films is similar to the 0.35 μV / K value observed in single-crystal Mn₃Sn [3]. Also shown in Fig. 4(b) is the $S_{xz}$ curve for the Fe reference device. The measured saturation $S_{xz}$ = 0.40 μV / K for Fe is in good agreement with that reported in the literature [28]. In ferromagnets, ANE is generally proportional to the magnetization. It is remarkable that ANE in antiferromagnetic Mn$_3$Ge is comparable to that of the strong ferromagnet Fe, even though Mn$_3$Ge has a ferromagnetic component of the magnetization that is three orders of magnitude lower.

### 4. Inverse spin Hall effect

Owing to the large spin-Hall conductivity that has been theoretically predicted for Mn$_3$Ge [13], a strong spin-to-charge conversion may be expected when a spin current is injected into Mn$_3$Ge. We employed ferromagnetic resonance-spin pumping (FMR-SP) to measure ISHE in our Mn$_3$Ge films. Fig. 5(a) displays the measured voltage ($V_{sp}$) from a Ru(10nm)/Mn$_3$Ge (100 nm)/ Py(10 nm) device at room temperature as a function of the applied field, along with the responses from the Pt/Py and Py reference devices measured under identical conditions.

The measured $V_{sp}$ is composed of a symmetric Lorentzian component ($V_{ISHE} = v_{ISHE}[\Delta H^2/(\Delta H^2 + (H - H_{FMR})^2)]$) due to the spin pumping-induced ISHE, and an antisymmetric Lorentzian component ($V_{AHE} = v_{AHE}[\Delta H(H - H_{FMR})/(\Delta H^2 + (H - H_{FMR})^2)]$) due to the rectified AHE voltage arising from the capacitive coupling-induced rf current and the magnetization precession in Py, along with a constant offset [29, 30]. The solid lines in Fig. 5(a) are the best-fit curves of the measured $V_{sp}$ for all three devices. The fitting routine allows us to extract the amplitudes of the ISHE and AHE contributions ($v_{ISHE}$ and $v_{AHE}$), along with the resonance field ($H_{FMR}$) and the linewidth ($\Delta H$) at all input rf frequencies. $v_{ISHE}$ is linearly proportional to the power of the applied rf excitation, indicating that the induced magnetization precession remains in the small-angle regime and that sample heating is negligible (See Supplemental Material [20] for detailed discussion). Fig. 5(b) shows the ISHE charge current ($V_{ISHE}/R$) for all three devices, where $R$ is the total device resistance measured across the contact pads using the four-probe method, $R_{Py} = 13\,\Omega$, $R_{Pt/Py} = 5.1\,\Omega$, $R_{Ru/Mn_3Ge/Py} = 3.2\,\Omega$. $V_{ISHE}/R$ of the Py single-layer device is negligible because, as expected, there is no ISHE. (The fact that $V_{sp}$ in the Py single-layer device is antisymmetric further rules out the presence of any Nernst-like signal due to heating effects in our measurements.) On the other hand, the peak $V_{ISHE}/R$ value of the Mn3Ge/Py device is significantly larger than that of the Pt/Py device. In view of the large variations in the numerical values of parameters related to spin transport and spin-to-charge conversion in the literature [32], we do not attempt to quantify the spin-Hall angle for our epitaxial Mn3Ge thin films. Instead, a direct comparison between Mn3Ge and Pt would be more intuitive.

By fitting the field dependence of the FMR frequency ($f_{res}$) using the Kittel formula, we have determined for the Py layers in our devices the gyromagnetic ratio $\gamma = 1.82 \times 10^{11}\,T^{-1}s^{-1}$, and the saturation magnetization $\mu_0 M_s = 0.84\,T$. These values are comparable to the ones reported in the literature [29-31]. In the inset of Fig. 5(b), the FMR linewidth ($\Delta H$), obtained from curve-fitting the $V_{sp}$ data, is plotted against $f_{res}$. Fitting the data points with the linear relation $\Delta H = (2\pi \alpha_{eff}/\gamma)f_{res} + \Delta H_0$ for all devices yields the Gilbert damping parameter $\alpha_{eff}$ for the Py, Pt/Py and Ru/Mn3Ge/Py devices as $0.0090 \pm 0.0004$, $0.0150 \pm 0.0008$ and $0.0300 \pm 0.0018$, respectively. The enhanced damping in the Pt/Py and Ru/Mn3Ge/Py devices is the result of spin pumping and is related to the spin mixing conductance $g_r^{\uparrow\downarrow}$ via

$$\alpha_{eff}^{spin-pumping} - \alpha_{eff}^{Py} = \left(\frac{\gamma \hbar}{4\pi M_s t_{Py}}\right)g_r^{\uparrow\downarrow}. \qquad (2)$$

Using the fitted values for $\alpha_{eff}$, $M_S$ and $\gamma$, we obtain $g_r^{\uparrow\downarrow}= 25.5 \pm 1.7$ nm$^{-2}$ for the Pt/Py interface, and $g_r^{\uparrow\downarrow}=90.5 \pm 4.9$ nm$^{-2}$ for Mn$_3$Ge/Py interface.

In a spin-pumping device, the ISHE voltage ($v_{ISHE}$) depends on the material and device parameters via

$$v_{ISHE} = \left(\frac{-e\theta_{SH}}{\sigma_{NM}t_{NM}+\sigma_{FM}t_{FM}}\right) \lambda \tanh\left(\frac{t_{NM}}{2\lambda}\right) g_r^{\uparrow\downarrow} fLP \left(\frac{\gamma h_{rf}}{2\alpha_{eff}\omega}\right)^2, \quad (3)$$

where $e$ is the electron charge, $\theta_{SH}$ is the spin Hall angle, $\sigma_{NM}$ ($\sigma_{FM}$) is the conductivity of NM (FM), $t_{NM}$ ($t_{FM}$) is the thickness of NM (FM), $\lambda$ is the spin diffusion length, $L$ is the sample length) [33]. The rf magnetic field $h_{rf}$ and the ellipticity of magnetization precession $P$ are unknown in our measurements. However, since our devices are identically fabricated, they are expected to have very similar impedance. Thus the $h_{rf}$ and $P$ terms cancel out when we take the ratio of $v_{ISHE}$ of devices measured under the same input rf power and frequency. The ratio of spin-Hall angles in Mn$_3$Ge and Pt is then give by

$$\frac{\theta_{SH}^{Mn_3Ge}}{\theta_{SH}^{Pt}} = \frac{(v_{ISHE}/R)_{Mn_3Ge}}{(v_{ISHE}/R)_{Pt}} * \frac{\lambda_{Pt}}{\lambda_{Mn_3Ge}} * \frac{\tanh\left(\frac{t_{Pt}}{2\lambda_{Pt}}\right)}{\tanh\left(\frac{t_{Mn_3Ge}}{2\lambda_{Mn_3Ge}}\right)} * \frac{(g_r^{\uparrow\downarrow})_{Pt}}{(g_r^{\uparrow\downarrow})_{Mn_3Ge}} * \left(\frac{\alpha_{eff}^{Mn_3Ge}}{\alpha_{eff}^{Pt}}\right)^2. \quad (4)$$

Using $\lambda_{Pt} \approx 3$ nm [31], and $\lambda_{Mn_3Ge} \approx 1$ nm [34], we estimate $\theta_{SH}^{Mn_3Ge}/\theta_{SH}^{Pt}$ to be $8 \pm 2$. The higher spin-mixing conductivity at the Mn$_3$Ge/Py interface and the larger spin-Hall angle in Mn$_3$Ge are consistent with the theoretical prediction of a large spin-Hall conductivity in Mn$_3$Ge [13].

### IV. CONCLUSIONS

In summary, we have synthesized continuous epitaxial thin films of the Kagome semimetal Mn$_3$Ge by magnetron sputtering. Large anomalous Nernst and inverse spin-Hall effects have been observed in thermoelectric and spin pumping devices from these films. Synthesis of crystallographically well-defined $Mn_3Ge$ thin films is an important step toward pursuing antiferromagnetic spintronics as well as elucidating the fundamental physics of topological materials.

*Note added: During the preparation of this manuscript, growth of continuous and epitaxial $Mn_3Sn$ film has been reported [35].*

## ACKNOWLEDGEMENT

We acknowledge Dr. Jonathan Gibbons for fruitful discussion in FMR measurement and Dr. Timothy Spila for the acquisition of RBS data. Work at Argonne is supported by the Center for the Advancement of Topological Semimetals, an Energy Frontier Research Center funded by the U.S. DOE, Office of Basic Energy Sciences. The use of facilities at the Center for Nanoscale Materials was supported by the U.S. DOE, BES under Contract No. DE-AC02-06CH11357.

\* Corresponding author: jiang@anl.gov

**References:**

[1] S. Nakatsuji, N. Kiyohara & T. Higo, Nature 527, 212 (2015).

[2] K. Kuroda, T. Tomita, M. -T. Suzuki, C. Bareille, A. A. Nugroho, P. Goswami, M. Ochi, M. Ikhlas, M. Nakayama, S. Akebi, R. Noguchi, R. Ishii, N. Inami, K. Ono, H. Kumigashira, A. Varykhalov, T. Muro, T. Koretsune, R. Arita, S. Shin, Takeshi Kondo and S. Nakatsuji, Nat. Mater. 16, 1090 (2017).

[3] M. Ikhlas, T. Tomita, T. Koretsune, M. -T. Suzuki, D. N. -Hamane, R. Arita, Y. Otani and S. Nakatsuji, Nat. Phys. 13, 1085 (2017).

[4] H. Yang, Y. Sun, Y. Zhang, W. -J. Shi, S. SP Parkin and B. Yan, New J. Phys. 19, 015008 (2017).

[5] J. Liu and L. Balents, Phys. Rev. Lett. 119, 087202 (2017).

[6] A. K. Nayak, J. E. Fischer, Y. Sun, B. Yan, J. Karel, A. C. Komarek, C. Shekhar, Ni. Kumar, W. Schnelle, J. Kubler, C. Felser and S. S. P. Parkin, Sci. Adv. 2, e1501870 (2016).

[7] L. Ye, M. Kang, J. Liu, F. von Cube, C. R. Wicker, T. Suzuki, C. Jozwiak, A. Bostwick, E. Rotenberg, D. C. Bell, L. Fu, R. Comin, and J. G. Checkelsky, Nature 555, 638 (2018).

[8] J.-X. Yin, S. S. Zhang, H. Li, K. Jiang, G. Chang, B. Zhang, B. Lian, C. Xiang, I. Belopolski, H. Zheng, T. A. Cochran, S.-Y. Xu, G. Bian, K. Liu, T.-R. Chang, H. Lin, Z.-Y. Lu, Z. Wang, S. Jia, W. Wang, and M. Z. Hasan, Nature 562, 91 (2018).

[9] L. Ye, M. K. Chan, R. D. Mcdonald, D. Graf, M. Kang, J. Liu, T. Suzuki, R. Comin, L. Fu and J. Checkelsky, Nat. Comm. 10, 4870 (2019).

[10] E. Liu, Y. Sun, N. Kumar, L. Muechler, A. Sun, L. Jiao, S. -Y. Yang, D. Liu, A. Liang, Q. Xu, J. Kroder, V. Süß, H. Borrmann, C. Shekhar, Z. Wang, C. Xi, W. Wang, W. Schnelle, S. Wirth, Y. Chen, S. T. B. Goennenwein and C. Felser, Nat. Phys. 14, 1125 (2018).

[11] J. -X. Yin, S. S. Zhang, G. Chang, Q. Wang, S. S. Tsirkin, Z. Guguchia, B. Lian, H. Zhou, K. Jiang, I. Belopolski, N. Shumiya, D. Multer, M. Litskevich, T. A. Cochran, H. Lin, Z. Wang, T. Neupert. S. Jia, H. Lei and M. Z. Hasan, Nat. Phys. 15, 443 (2019).

[12] J. Kübler and C. Felser, EPL 120, 47002 (2017).

[13] Y. Zhang, Y. Sun, H. Yang, J. Zelezny, S. P. P. Parkin, C. Felser and B. Yan, Phys. Rev. B 95, 075128 (2017).

[14] M. Kimata, H. Chen, K. Kondou, S. Sugimoto, P. K. Muduli, M. Ikhlas, Y. Omori, T. Tomita, A. H. MacDonald, S. Nakatsuji and Y. Otani, Nature 565, 627 (2019).

[15] T. Higo, H. Man, D. B. Gopman, L. Wu, T. Koretsune, O. M. J. van't Erve, Y. P. Kabanov, D. Rees, Y. Li, M. -T. Suzuki, S. Patankar, M. Ikhlas, C. L. Chien, R. Arita, R. D. Shull, J. Orenstein and S. Nakatsuji, Nat. Photon. 12, 73 (2018).

[16] N. Kiyohara, T. Tomita, and S. Nakatsuji, Phys. Rev. Applied 5, 064009 (2016).

[17] T. Ogasawara, J. Kim, Y. Ando, A. Hirohata, J. Magn. Magn. Mater. 473, 7 (2019).

[18] A. Markou, J. M. Taylor, A. Kalache, P. Werner, S. S. P. Parkin and C. Felser, Phys. Rev. Mater. 2, 051001 (R) (2018).

[19] M. Björck and G. Andersson J. Appl. Cryst. 40, 1174 (2007).

[20] See Supplemental Material for further details on experimental measurements of the $Mn_3Ge$ films.

[21] S. Yamada, Y. Nishibe, M. Saizaki, H. Kitajima, S. Ohtsubo, A. Morimoto, T. Shimizu, K. Ishida, and Y. Masaki, Jpn. J. Appl. Phys. 41, L206 (2002).

[22] ICSD database, https://icsd.fiz-karlsruhe.de/index.xhtml.

[23] S. Tomiyoshi, Y. Yamaguchi, and T. Nagamiya, J. Magn. Magn. Mater. 31–34, 629 (1983).

[24] J. Noky, J. Goth, C. Felser, and Y. Sun, Phys. Rev. B 98, 241106 (R) (2018).

[25] C. Wuttke, F. Caglieris, S. Sykora, F. Scaravaggi, A. U. B. Wolter, K. Manna, V. Süss, C. Shekhar, C. Felser, B. Büchnerk and C. Hess, Phys. Rev. B 100, 085111 (2019).

[26] Y. Song, Y. Qiao, Q. Huang, C. Wang, X. Liu, Q. Li, J. Chen, and X. Xing, Chem. Mater. 30, 6236 (2018).

[27] L. Xu, X. Li, X. Lu, C. Collignon, H. Fu, J. Koo, B. Fauque, B. Yan, Z. Zhu, K. Behnia, Sci. Adv. 6, eaaz3522 (2020).

[28] J. Weischenberg, F. Freimuth, S. Blugel, and Y. Mokrousov, Phys. Rev. B 87, 060406(R) (2013).

[29] K. Ando, S. Takahashi, J. Ieda, Y. Kajiwara, H. Nakayama, T. Yoshino, K. Harii, Y. Fujikawa, M. Matsuo, S. Maekawa, and E. Saitoh, J. Appl. Phys. 109, 103913 (2011).

[30] Jing Zhou, Xiao Wang, Yaohua Liu, Jihang Yu, Huixia Fu, Liang Liu, Shaohai Chen, Jinyu Deng, Weinan Lin, Xinyu Shu, Herng Yau Yoong, Tao Hong, Masaaki Matsuda, Ping Yang, Stefan Adams, Binghai Yan, Xiufeng Han and Jingsheng Chen, Sci. Adv. 5, eaau6696 (2019).

[31] L. Liu, T. Moriyama, D. C. Palph, and R. A. Buhrman, Phys. Rev. Lett. 106, 036601 (2011).

[32] J. Sinova, S.O. Valenzuela, J. Wunderlich, C.H. Back, and T. Jungwirth, Rev. Mod. Phys. 87, 1213 (2015).

[33] H. L. Wang, C. H. Du, Y. Pu, R. Adur, P. C. Hammel, and F. Y. Yang, Phys. Rev. Lett. 112, 197201 (2014).

[34] P. K. Muduli, T. Higo, T. Nishikawa, D. Qu, H. Isshiki, K. Kondou, D. Nishio-Hamane, S. Nakatsuji, and YoshiChika Otani, Phys. Rev. B 99, 184425 (2019).

[35] J. M. Taylor, A. Markou, E. Lesne, P. K. Sivakumar, C. Luo, F. Radu, P. Werner, C. Felser, and S. S. P. Parkin, Phys. Rev. B 101, 094404 (2020).

**Supplemental Material reference：**

[36] J.F. Qian, A.K. Nayak, G. Kreiner, W. Schnelle, and C. Felser, J. Phys. D: Appl. Phys. 47, 305001 (2014).

**Fig. 1**

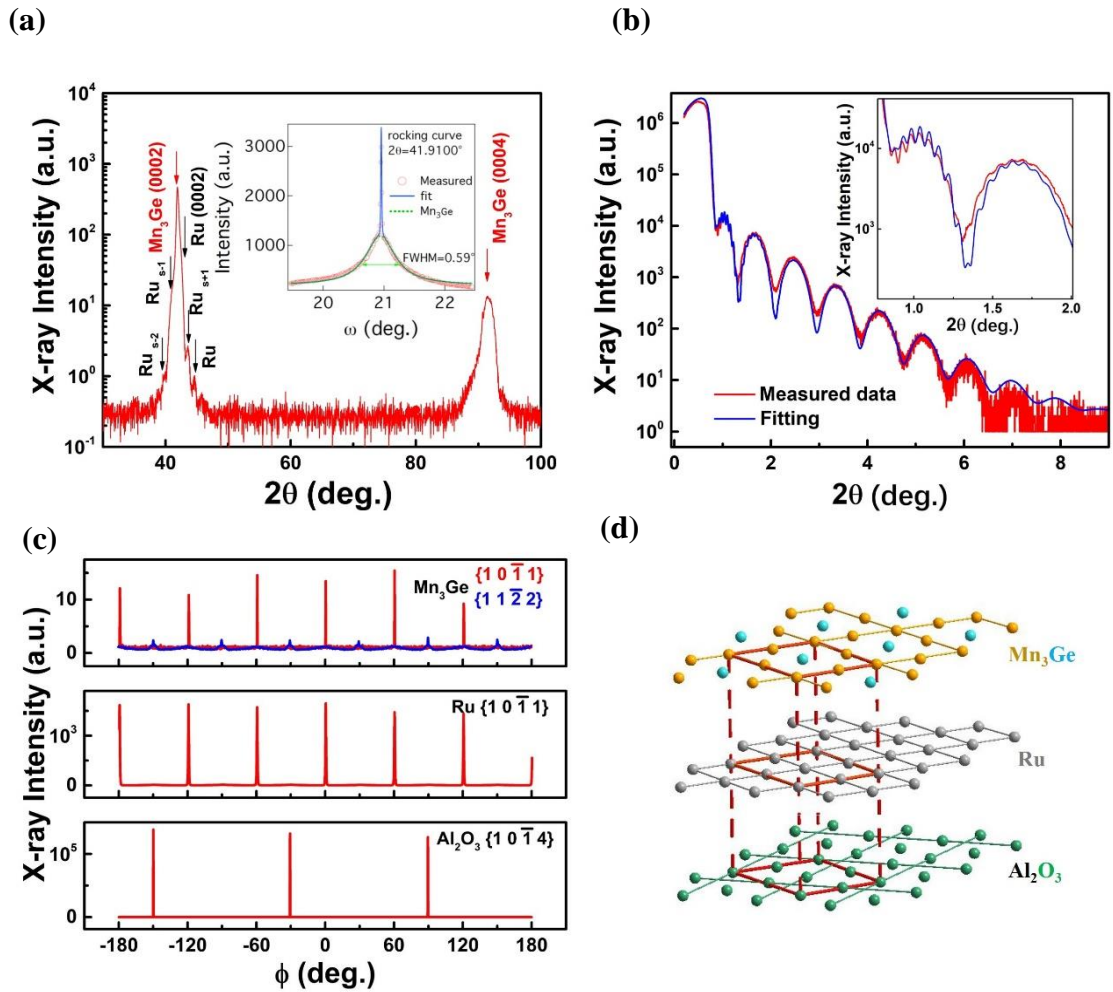

Fig. 1. (a) X-ray diffraction pattern of a sputter-grown sapphire/Ru(10nm)/Mn$_3$Ge(100nm) film. Ru satellite peaks are labeled as s followed by their order. Inset: rocking curve of Mn$_3$Ge (0002) at $2\theta = 41.91°$ (The blue solid curve is a fit and the green dotted curve is the Mn$_3$Ge contribution). (b) X-ray reflectivity curve of the same film. The red and blue curves are the measured data and simulation, respectively. Inset: a zoomed-in view showing details of the long-period oscillations due to the Ru layer and the short-period oscillations due to the Mn$_3$Ge layer. (c) Phi scans around partially in-plane peaks (as labeled) of Mn$_3$Ge, Ru and sapphire. (d) Stacking configuration among Mn$_3$Ge, Ru and sapphire according to the phi scans. The vertical red dashed lines illustrate atomic registry. For simplicity, oxygen termination in the sapphire substrate is chosen.

**Fig. 2**

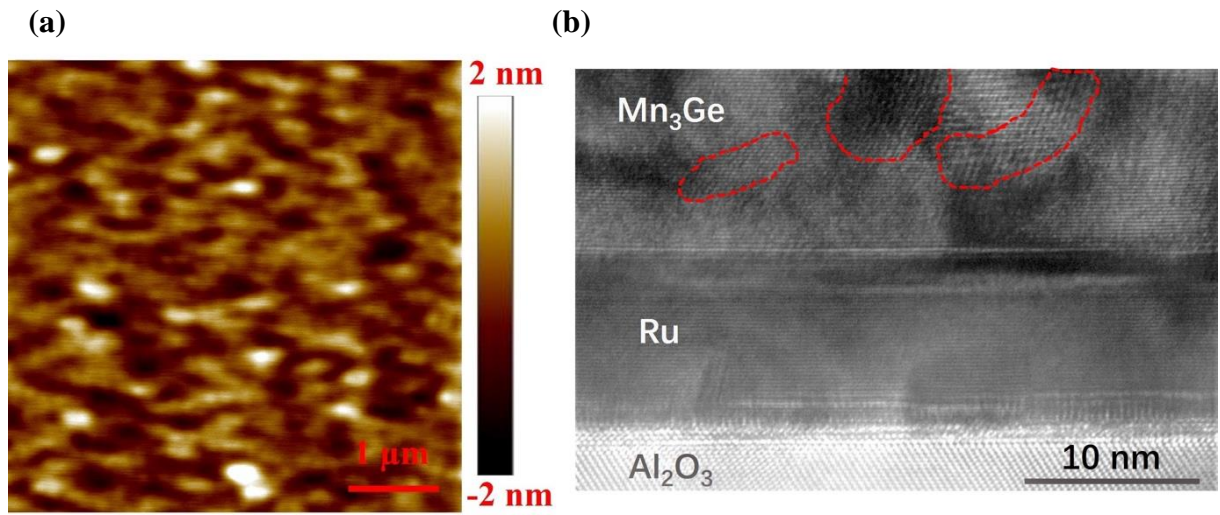

Fig. 2. (a) AFM image of a sapphire/Ru(10nm)/Mn$_3$Ge(100nm) film showing an RMS roughness of ~ 2.5 nm over the 5 x 5 μm$^2$ area. (b) Cross-sectional HRTEM image viewed along the sapphire $[10\bar{1}0]$ axis. Areas with Moiré fringes are outlined in red.

**Fig. 3**

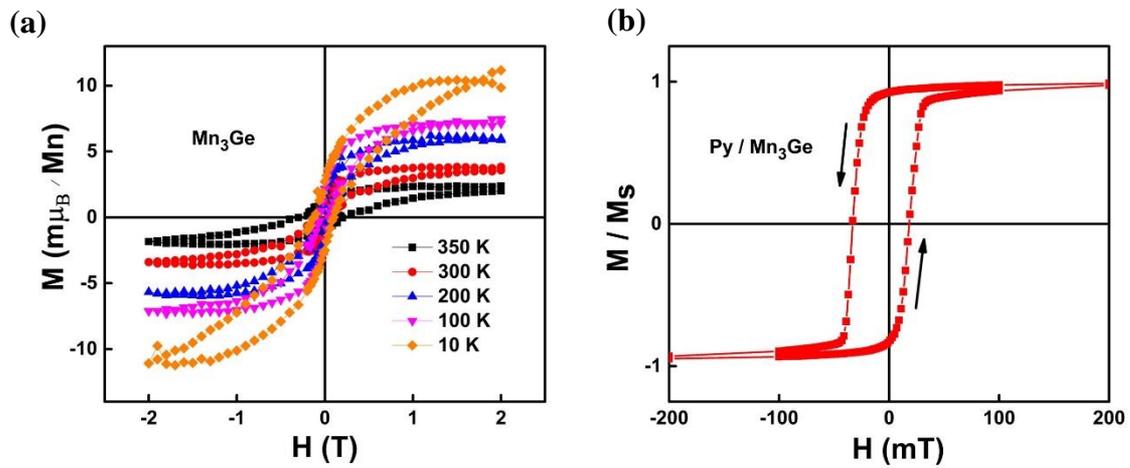

Fig. 3. (a) The ferromagnetic component of the in-plane magnetization in a 100 nm $Mn_3Ge$ film measured at various temperatures. (b) The hysteresis loop of a $Mn_3Ge$(100 nm)/Py(10 nm) exchange-bias structure measured at 10 K, after field-cooling in 1 T from 300 K.

**Fig. 4**

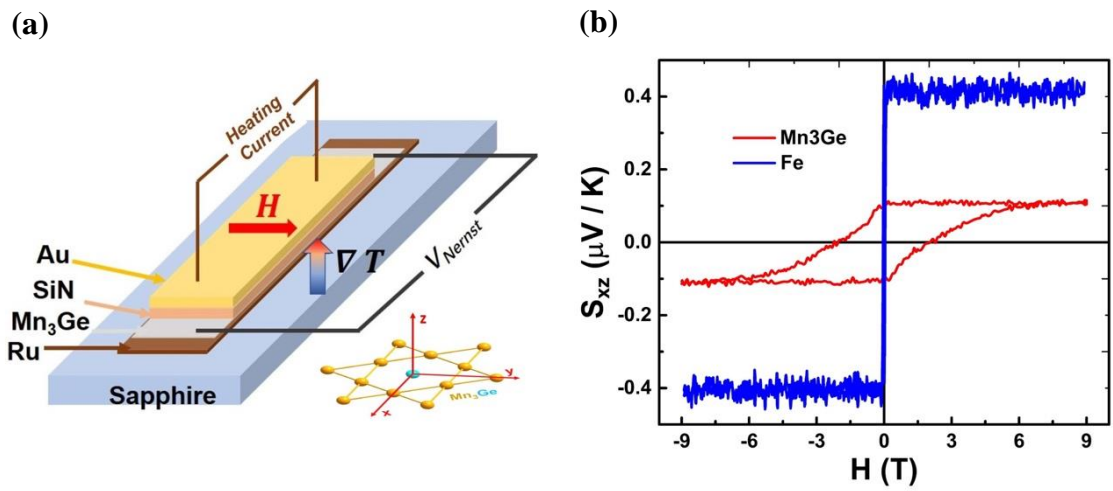

Fig. 4. (a) Schematic of the Nernst effect device. (b) Field dependence of the ANE signal $S_{xz}$ of Mn$_3$Ge and Fe measured at 300 K.

**Fig. 5**

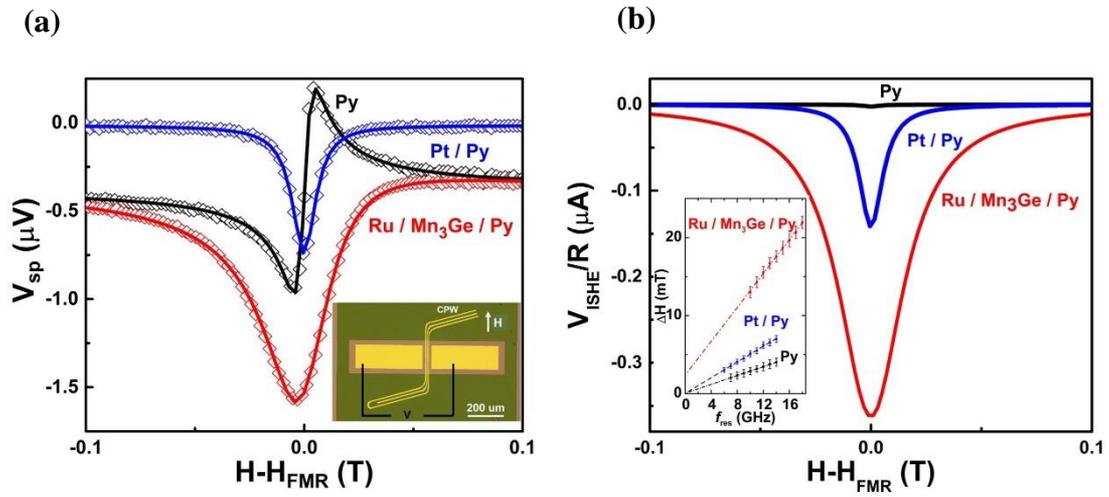

Fig. 5. (a) FMR-spin pumping voltage $V_{sp}$ plotted as a function of the applied field in Py, Pt/Py, Ru/Mn3Ge/Py devices measured at 13 GHz and 18 dBm at room temperature. The solid lines are fits to the measured data. Inset: Optical image of a spin-pumping device. (b) The ISHE current $V_{ISHE}/R$ in the devices extracted from curve fitting. Inset: FMR linewidth ($\Delta H$) plotted as a function of the resonance frequency. The dashed lines are linear fits.